# Electrically controlled emission from triplet charged excitons in atomically thin heterostructures

Andrew Y. Joe[1], Luis A. Jauregui[2], Kateryna Pistunova[1], Zhengguang Lu[3,4], Dominik S. Wild[1], Giovanni Scuri[1], Kristiaan De Greve[1,5+], Ryan J. Gelly[1], You Zhou[1,5], Jiho Sung[1,5], Andrés Mier Valdivia[6], Andrey Sushko[1], Takashi Taniguchi[7], Kenji Watanabe[7], Dmitry Smirnov[3], Mikhail D. Lukin[1], Hongkun Park[1,5] and Philip Kim[1,6]*

[1] *Department of Physics, Harvard University, Cambridge, Massachusetts 02138, USA*
[2] *Department of Physics and Astronomy, University of California, Irvine, California 92697, USA*
[3] *National High Magnetic Field Laboratory, Tallahassee, Florida 32310, USA*
[4] *Department of Physics, Florida State University, Tallahassee, Florida 32306, USA*
[5] *Department of Chemistry and Chemical Biology, Harvard University, Cambridge, Massachusetts 02138, USA*
[6] *John A. Paulson School of Engineering and Applied Sciences, Harvard University, Cambridge, Massachusetts 02138, USA*
[7] *National Institute for Materials Science, 1-1 Namiki, Tsukuba 305-0044, Japan*
+ *currently at IMEC, 3001 Leuven, Belgium*

\* *To whom correspondence should be addressed: pkim@physics.harvard.edu*

**Excitons are composite bosons that can feature spin singlet and triplet states. In usual semiconductors, without an additional spin-flip mechanism, triplet excitons are extremely inefficient optical emitters. Large spin-orbit coupling in transition metal dichalcogenides (TMDs) couples circularly polarized light to excitons with selective valley and spin[1–4]. Here, we demonstrate electrically controlled brightening of spin-triplet interlayer excitons in a MoSe$_2$/WSe$_2$ TMD van der Waals (vdW) heterostructure. The atomic registry of vdW layers in TMD heterostructures provides a quasi-angular momentum to interlayer excitons[5,6], enabling emission from otherwise dark spin-triplet excitons. Employing magnetic field, we show that photons emitted by triplet and singlet excitons in the same valley have opposite chirality. We also measure effective exciton *g*-factors, presenting direct and quantitative evidence of triplet interlayer excitons. We further demonstrate gate tuning of the relative photoluminescence intensity between singlet and triplet charged excitons. Electrically controlled emission between singlet and triplet excitons enables a route for optoelectronic devices that can configure excitonic chiral, spin, and valley quantum states.**

Inefficient light emission from triplet excitons is known to be a major bottleneck for many optoelectronic devices, such as e.g. organic LEDs, where a significant portion of randomly generated excitons are in spin-triplet states[7]. The search for highly emissive "bright" triplet excitons generally involves materials with strong spin-orbit coupling (SOC)[8] that also provide such spin-flip mechanisms with altered optical selection rules. Electronic band structure engineering in van der Waals (vdW) heterostructures of transition metal dichalcogenides (TMDs) with strong SOC may produce a unique material platform to realize emissive triplet excitons.

Semiconducting TMDs exhibit extraordinary excitonic effects when reduced to the two-dimensional limit[9–11]. Monolayer TMDs have large exciton binding energies[12] and spin-valley locking[13,14], which can be harnessed for optoelectronic[15,16] and valleytronic[14] applications. When monolayer TMDs are stacked together to form heterobilayers such as WSe$_2$/MoSe$_2$[1–4,17–21], MoSe$_2$/MoS$_2$[22,23], or WS$_2$/MoS$_2$[24–26], interlayer excitons (IEs) can form across the atomically sharp interfaces owing to their type-II band alignment[27] and ultrafast charge transfer[22,24,25] between the layers. The resulting IEs have long lifetimes[17,19,20], a permanent out-of-plane dipole moment[3,19], and modified optical selection rules[2–6] due to the electrons and holes residing in separate layers. When the heterostructures are electron or hole doped, the IEs bind with free carriers to form charged interlayer excitons (CIEs)[19].

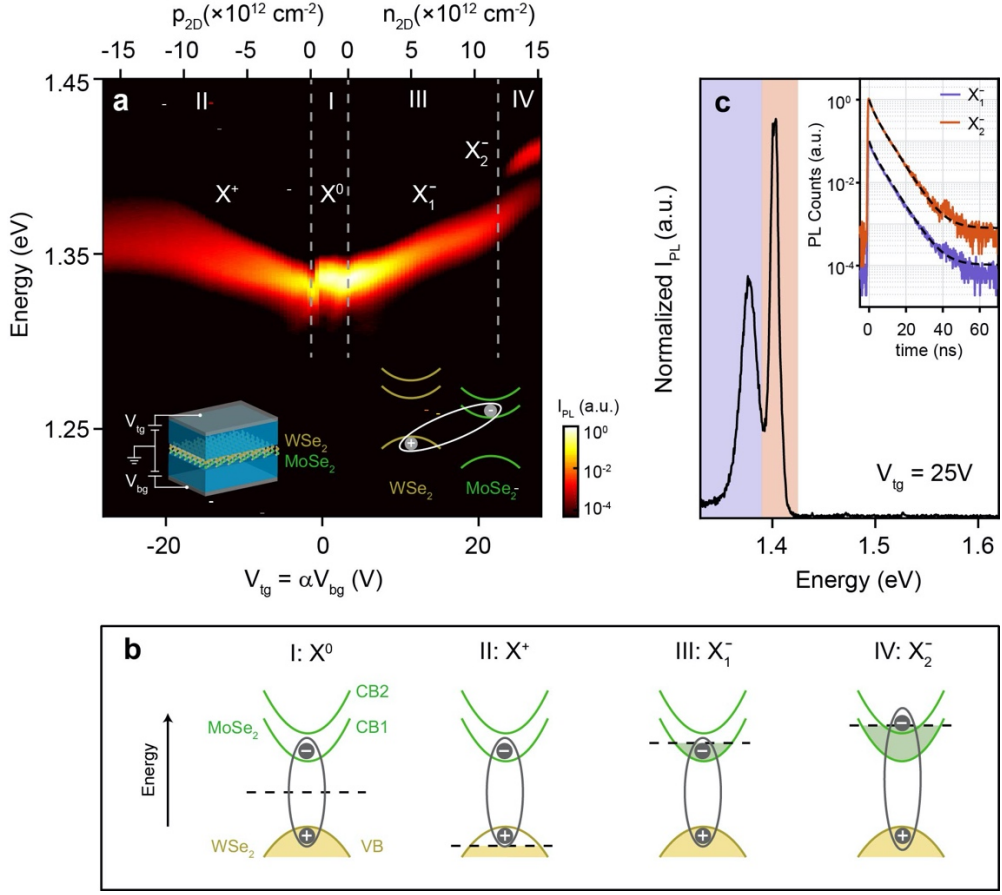

**Figure 1 | Emerging photoluminescence spectra at large *n*-doped regime. a,** PL vs. $V_{tg} = \alpha V_{bg}$, where $\alpha = 0.617$ based on the top and bottom *h*-BN thicknesses. We electrostatically tune the doping through regions I-IV (separated by gray dashed lines), observing emission from neutral interlayer excitons ($X^0$), *p*-doped charged excitons ($X^+$), *n*-doped charged excitons ($X_1^-$), and a new higher energy peak ($X_2^-$). Left inset: device schematic and direction of applied gate voltages. Right inset: band schematic of an interlayer exciton between the conduction band of MoSe$_2$ and valence band of WSe$_2$. **b,** Reduced band diagrams of the MoSe$_2$/WSe$_2$ heterostructure showing the upper (CB2) and lower (CB1) conduction bands of MoSe$_2$ and the valence band of WSe$_2$ (VB). The exciton and Fermi energy (black dashed line) is drawn for each regime marked in **(a)**. The green (MoSe$_2$) and yellow (WSe$_2$) shaded areas indicate filled electron bands. **c,** PL spectra at $V_{tg} = 25$ V with the $X_1^-$ and $X_2^-$ regions shaded in purple and orange, respectively. Inset: time-dependent PL measurements to determine the lifetime of $X_1^-$ ($\tau_1 \sim 6.12$ ns) and $X_2^-$ ($\tau_2 \sim 6.08$ ns) by spectrally filtering the energy range indicated in **(c)**. Dashed black lines are exponential fits.

Our experiments employ *h*-BN encapsulated WSe$_2$/MoSe$_2$ devices with top and bottom gates, and electrically transparent contacts (Fig. 1a left inset and Supplementary Fig. 1), as described in the previous work[19]. We use a dual-gating scheme where the top-gate voltage ($V_{tg}$) and the back-gate voltage ($V_{bg}$) have the same polarity, achieving higher carrier densities than in previous IE studies[3,19,20] (details in Supplementary Section 1).

Figure 1a shows the photoluminescence (PL) spectrum as a function of $V_{tg} = \alpha V_{bg}$, where $\alpha = 0.617$ (based on top and bottom *h*-BN thicknesses measured via atomic force microscopy). We identify four distinct gate regions, marked by I-IV, based on the electrostatic doping of the heterostructure (Fig. 1b). We verify the doping of the layers by measuring the intralayer exciton absorption spectra as a function of the gate voltage (Supplementary Fig. 2). In region I, we observe neutral interlayer excitons labeled as $X^0$. In region II (III), the Fermi energy crosses the valence band of WSe$_2$ (lower conduction band (CB1) of MoSe$_2$) and we begin to *p*-dope (*n*-dope) the heterostructure forming CIEs, $X^+$ ($X_1^-$). The discontinuities in the PL energy between regions I/II and I/III are attributed to CIE binding energies of ~ 15 meV and ~ 10 meV, respectively, similar to our previous work[19]. Surprisingly, in region IV, when the electron density ($n_{2D}$) is further increased, an additional PL peak, $X_2^-$, appears ~ 25 meV above the $X_1^-$ peak, which overtakes in intensity with increasing $n_{2D}$. This additional exciton feature is likely related to reaching the upper conduction band of MoSe$_2$ (CB2), as shown in Figure 1b.

The appearance of spectrally resolved exciton branches in region IV enables us to measure the lifetimes of these excitons. We spectrally isolate the two peaks obtained at $V_{tg} = \alpha V_{bg} = 25$ V (Fig. 1c) and perform time-dependent PL (TDPL) measurements to find the lifetimes corresponding to $X_1^-$ and $X_2^-$ are $\tau_1 = 6.08 \pm 0.01$ ns and $\tau_2 = 6.12 \pm 0.02$ ns (Fig. 1c inset), respectively. The measured lifetimes are similar for the two species, but still 3 to 4 orders of magnitude longer than the typical lifetime of neutral or charged excitons in monolayer TMDs (~ 1 ps[28,29]).

Since the excitons in TMDs are formed at two valleys, +K and –K, at the corners of the Brillouin zone, their magnetic moment contains contributions from angular momenta of valley, spin, and orbital components of the exciton wavefunction. To understand the angular momentum characteristics of the interlayer excitons observed in our experiments, we thus measure PL under magnetic fields to determine the effective Zeeman splitting of the exciton species. We perform polarization-resolved PL measurements as a function of magnetic field (*B*) using a cross-polarized measurement scheme (Supplementary Fig. 3a).

Figures 2a-c show the normalized σ+ (blue) and σ– (red) PL spectra at $V_{bg} = \alpha V_{tg} = 0$, 17, and 25 V, respectively. From these polarization-resolved spectra, we obtain the PL energy splitting between the circularly polarized light ($\Delta E = E_{\sigma+} - E_{\sigma-}$) as a function of *B*. Figure 2d shows the measured energy difference follows a linear relation $\Delta E = g\mu_B B$, where *g* is the effective *g*-factor and $\mu_B$ is the Bohr magneton. From the slope of measured relation between $\Delta E$ and *B*, we obtain the effective *g*-factors for $X^0$, $X_1^-$, and $X_2^-$: $g_0 = 6.99 \pm 0.35$, $g_1 = 6.06 \pm 0.58$, and $g_2 = -10.6 \pm 1.0$, respectively. Interestingly, the *g*-factor for $X_2^-$ is greater than and has the opposite sign of $g_0$ and $g_1$, implying $X_2^-$ has an additional Zeeman splitting contribution and that the chiral light coupling to the K valleys is flipped compared to $X^0$ or $X_1^-$.

To explain the emergence of $X_2^-$ and its unexpected *g*-factor, we consider the well-established band alignment diagram of a 0-degree aligned WSe$_2$/MoSe$_2$ heterostructure with spin-valley locking. The observation of a higher energy emission in region IV suggests that transitions between the highest WSe$_2$ K-valley valence band and both spin-split MoSe$_2$ K-valley conduction bands are allowed. This would indicate that the higher energy peak is an emissive triplet transition with an in-plane dipole moment, unlike dark triplet excitons in monolayers[30–32]. Modified selection rules arise from the space group symmetry of the atomic configuration of WSe$_2$/MoSe$_2$ heterostructures – as opposed to being purely determined by the symmetry of any individual, constituent layer. It was theoretically and experimentally shown that for WSe$_2$/MoSe$_2$ heterostructures, the lowest energy atomic stacking registry is $R_h^X$, representing an *R*-type stacking (0-degree aligned) with the *X* site of the top MoSe$_2$ layer above the *h* site of the bottom WSe$_2$ layer (Supplementary Fig. 4)[2,5,6]. In this stacking configuration, the optical selection rules for interlayer recombination involve the phase difference from the translation of the Bloch wave function between the two rotation sites, $h$ and $X$, in addition to the orbital angular momentum quantum number (Supplementary Section 4). This induced phase from the atomic registry translation can be regarded as an additional quasi-angular momentum in the optical transition, altering the resulting selection rules (Supplementary Table 1). These selection rules show that a transition from CB2 is optically allowed, consistent with our experimental observations. Moreover, they predict spin-singlet and spin-triplet IEs that have opposite circular light polarization coupling to the K valleys (Supplementary Fig. 5a).

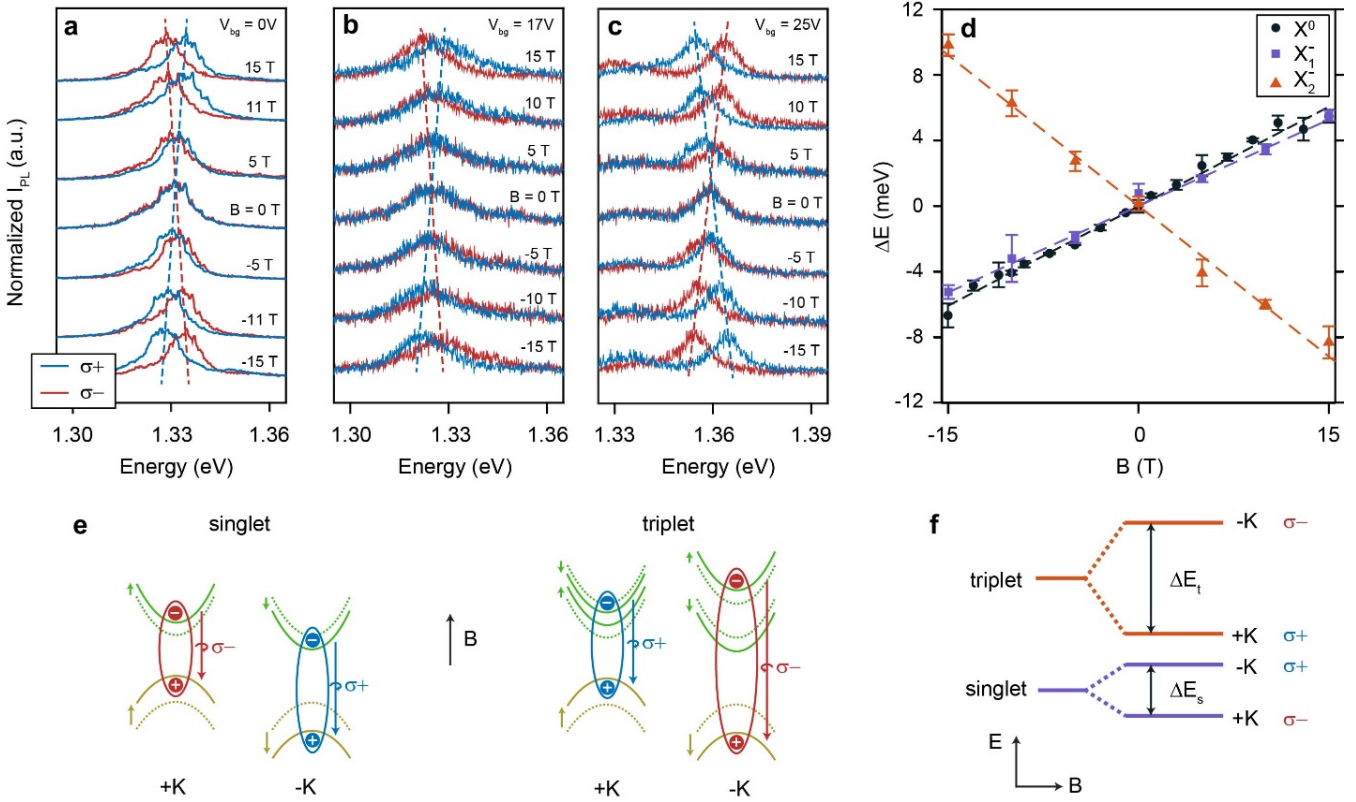

**Figure 2 | Experimental evidence of spin-singlet and spin-triplet excitons via magnetic field. a-c,** Polarization-resolved photoluminescence (PL) spectra at $V_{bg} = \alpha V_{tg} = 0$, 17, and 25 V, respectively, for characteristic magnetic fields ($B$) using a cross-polarized measurement scheme (Supplementary Section 3). Blue (red) curves are σ+ (σ−) collection. The dashed lines serve as a guide to the eye. **d,** PL energy splitting ($\Delta E = E_{\sigma+} - E_{\sigma-}$) as a function magnetic field for $X^0$, $X_1^-$, and $X_2^-$. Error bars are calculated from the fitting of the peak position. The dashed lines are linear fits to the energy splitting giving $g_0 = 6.99 \pm 0.35$, $g_1 = 6.06 \pm 0.58$, and $g_2 = -10.6 \pm 1.0$. **e,** Band diagrams and Zeeman splitting for the singlet and triplet transitions without (dashed lines) and with (solid lines) magnetic field. **f,** Zeeman splitting for the singlet and triplet excitons in the exciton particle picture and the circular polarization light coupling. The triplet exciton has an enhanced $g$-factor due to spin contributions.

Evidence for the singlet and triplet states and opposite circular polarization coupling is revealed by calculating the expected exciton $g$-factors. Using the single electron band picture[17,33], theoretically expected values for the $g$-factor can be determined based on the band alignment of the 0-degree aligned MoSe$_2$/WSe$_2$ heterostructure. The $g$-factor can be broken into the Zeeman shift of each electron band (Fig. 2e), which has contributions from the spin ($\mu_s$), the atomic orbitals ($\mu_l$), and the valley magnetic moment ($\mu_{\pm K}$) (further details in Supplementary Section 5). The exciton $g$-factor can then be determined as the relative shift of the conduction and valence bands for each circularly polarized transition. From these arguments, we calculate the singlet g-factor to be $g_{singlet}^{theory} \approx 7.1$, with contributions from only the atomic orbitals and the valley magnetic moments of the bands. The triplet g-factor, however, is $g_{triplet}^{theory} \approx -11.1$, with an additional contribution from the spins and the opposite sign due to the flipped circular polarized light coupling. These calculated g-factors are in excellent agreement with experimentally observed values both in terms of sign and magnitude, confirming spin singlet and triplet assignments for $X_1^-$, and $X_2^-$. We note that unlike the traditional picture of singlet and triplet states, the degeneracy of interlayer exciton singlet and triplet states is already broken due to spin-orbit coupling. These states split differently under magnetic field as shown in Figure 2f in the exciton particle picture. We also remark that the observed g-factors in our experiment are inconsistent with a 60-degree aligned sample (Supplementary Section 7). Thus, we confirm the $R_h^X$ stacking configuration for our heterostructure and demonstrate direct evidence of spin-singlet and triplet excitons.

From these measurements, we can now assign the peaks as either singlet or triplet states. In regions I-III, $X^0$, $X^+$, and $X_1^-$ all have transitions from CB1, allowing us to assign them as singlet neutral or singlet charged excitons. In region IV, the $X_2^-$ peak is a transition from CB2 in the presence of free carriers and is therefore a triplet charged exciton. The emergence of $X_2^-$ only after sufficient band filling can be explained by relative dipole strength, and PL being dominated by the lowest energy transition. Our lifetime measurements of $X_1^-$ and $X_2^-$ showed that $\tau_1 \approx \tau_2$, suggesting the optical dipole strength of the two exciton species are similar, consistent with theoretical calculations[5]. Therefore, although the triplet transition is allowed, we only observe $X_2^-$ once the inter-conduction band relaxation rates are quenched by Pauli blocking (further discussion in Supplementary Section 9).

Understanding the spin-state of the excitons allows us to control the optical behavior of charged excitons in regions II and III under magnetic field. Figures 3a-b show polarization-resolved PL measurements at $B = 15$ T in the band filling regimes corresponding to p-, neutral and n-doped singlet excitons. To minimize the induced vertical electric field, we apply $V_{bg}$ ($V_{tg}$) for negative (positive) voltages so that the electric field is screened and remains constant once the WSe$_2$ (MoSe$_2$) layer is doped[19]. Using the same cross-polarization measurement scheme described above, we observe the σ+ (σ−) emission intensity to be larger for the $X^+$ ($X_1^-$). Figure 3d shows the normalized difference in PL peak intensity, $\Delta PL_{norm} = (I_{PL}^{\sigma+} - I_{PL}^{\sigma-})/(I_{PL}^{\sigma+} + I_{PL}^{\sigma-})$, as a function of gate voltage for $B = -15$, 0, and 15 T. Without magnetic field, $\Delta PL_{norm}$ stays nearly constant, indicating excitons with opposite chirality are nearly degenerate. At $B = 15$ T, $\Delta PL_{norm}$ is positive (negative) for the $X^+$ ($X_1^-$), while the effect is reversed for $B = -15$ T.

Assuming $\Delta PL_{norm}$ is proportional to the percentage of charged exciton density imbalance towards σ+ circular polarization, our experimental observations imply a magnetic field induced imbalance in CIE density between the two valleys due to Zeeman splitting of the bands. CIEs ($X^+$ and $X_1^-$), at low doping, can only bind to a free carrier from the opposite valley due to the Pauli exclusion principle. Therefore, for the σ+ exciton, the hole (electron) must come from the +K valley (Fig. 3c). Taking $X^+$, the hole density is expected to be larger in the +K valley than in the −K valley for $B > 0$ (Fig. 3d left inset). Since the emitting exciton binds with a hole from the opposite valley, higher hole density in the +K valley induces higher emission of σ+ (−K valley) $X^+$ for $B > 0$, in agreement with our experimental results. Similarly, in the conduction band (Fig. 3d right inset), we expect a higher density of σ− for $X_1^-$ when $B > 0$, in agreement with the experiment.

Increasing gate voltage further into region IV under magnetic fields, we now can demonstrate gate tunable singlet-triplet CIE transitions inducing electrically tunable chiral optical response in our devices. Figures 4a-b shows the doping-dependent polarization-resolved PL in the singlet-triplet transition regime at $B = 15$ T. Here, we apply a dual-gating scheme $V_{bg} = \alpha\, V_{tg}$, reaching $n_{2D} \sim 10^{13}$ cm$^{-2}$, sweeping through regions III to IV in Fig. 1a. Figures 4c-d compare the $X_2^-$ and $X_1^-$ PL intensity as a function of the gate at $B$ = 0 and 15 T and we observe $X_2^-$ overtake $X_1^-$ in PL intensity with carrier density.

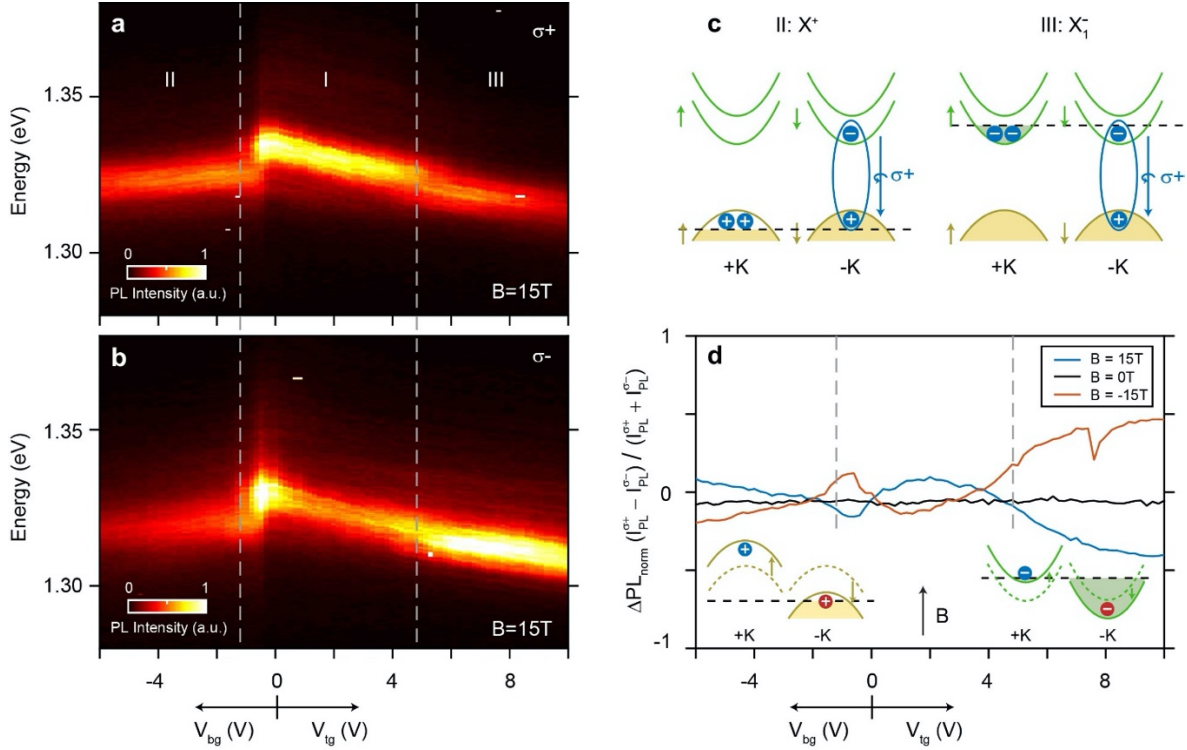

**Figure 3 | Charged nature of spin-singlet charged excitons under magnetic field. a,** Normalized photoluminescence (PL) spectra with σ+ collection and a cross-polarization measurement scheme at $B$ = 15 T vs $V_{tg}$ ($V_{bg}$) for positive (negative) voltages. The grey dashed lines separate the three regions. **b,** Same as **(a)** for σ− collection at $B$ = 15 T. **c,** Band diagrams of σ+ circularly polarized transitions for $X^+$ and $X_1^-$ showing the free carriers must come from the opposite valley. The dashed black line indicates the Fermi energy and the arrows indicate the band spin. **d,** $\Delta PL_{norm} = (I_{PL}^{\sigma+} - I_{PL}^{\sigma-})/(I_{PL}^{\sigma+} + I_{PL}^{\sigma-})$ as a function of gate voltage for $B$ = -15, 0, and 15 T. For $B$ = 15 T, $\Delta PL_{norm}$ is positive (negative) in the valence (conduction) band because the magnetic field splitting changes the hole (electron) density in the two valleys. Right inset: band diagram of the WSe₂ valence band without (dashed lines) and with (solid lines) magnetic field and the approximate location of the Fermi energy (dashed black line). Left inset: same diagram for the MoSe₂ conduction band.

Quantitative understanding of the gate dependent PL spectra can be made considering the band filling of both CB1 and CB2 as a function of $n_{2D}$, controlled by both gates. Figure 4c shows that the transition from $X_1^-$ to $X_2^-$ occurs at $n_{2D}^* \approx 1.3 \times 10^{13}$ cm⁻² (density calculated from capacitor model in Supplementary Section 1), where the PL emission is overtaken by $X_2^-$. Taking the effective mass of CB1 to be $m_e^{(1)} = 0.8\, m_0$[34] in units of bare electron mass ($m_0$), the $n_{2D}^*$ value obtained above allows us to estimate the energy separation between the spin-split MoSe₂ conduction band minima: $\Delta E_{CB} = \frac{2\pi n_{2D}^* \hbar^2}{\nu m_e^{(1)}} \approx 39$ meV, where we use $\nu = 2$ for the K-valley degeneracy factor. The estimated $\Delta E_{CB}$ value is consistent with previous experimental values[34], but larger than theoretically calculated values[32,35]. We note that we use the effective mass from recent transport studies[34] for all calculations, which suggest a systematic underestimate of the effective mass in DFT calculations[35].

Applying a magnetic field in the singlet-triplet transition regime, we demonstrate a flip in the dominant polarized light and tuning of the transition due to the Zeeman splitting of CB2. Figure 4d shows the polarization-resolved PL intensities for $X_1^-$ and $X_2^-$ across the transition at $B$ = 15 T. First, we find that the difference in polarized PL intensity is opposite for $X_1^-$ and $X_2^-$. In the band filling model mentioned above, the $X_2^-$ population is controlled by Pauli blocking rather than free carrier density. For $X_2^-$, when the

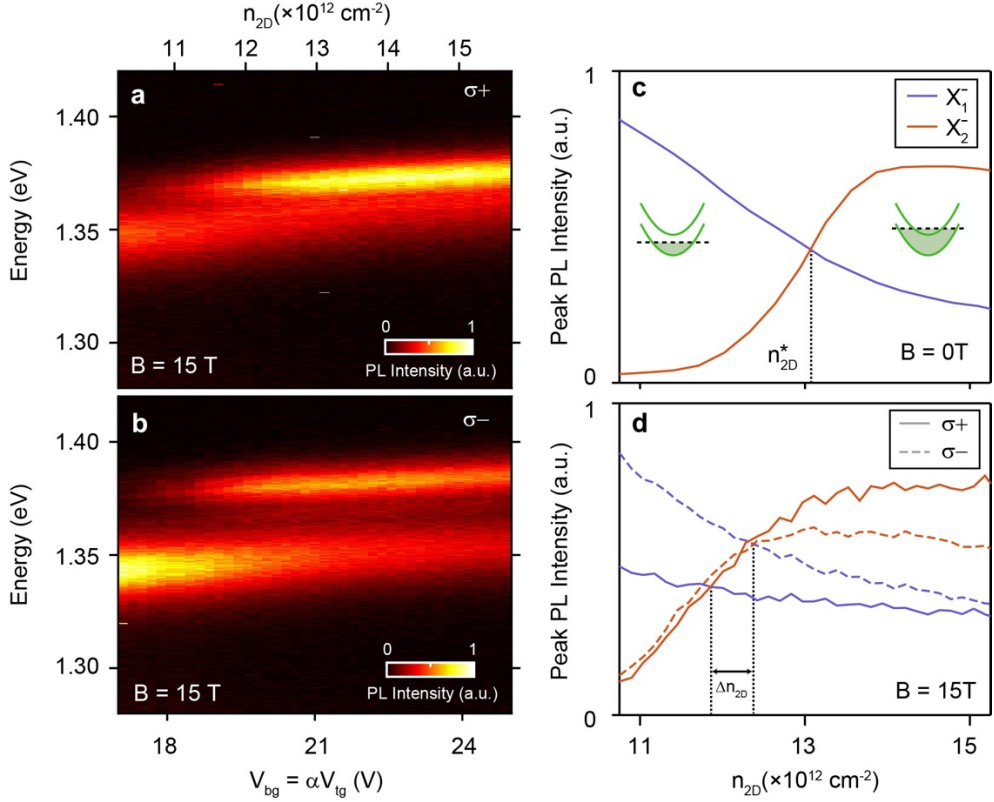

**Figure 4 | Tuning of the spin-triplet state under magnetic field. a,** Photoluminescence (PL) spectra with σ+ collection and a cross-polarization measurement scheme at $B$ = 15 T vs $V_{bg} = \alpha V_{tg}$. The corresponding carrier density is on the top x-axis. **b,** Same as (a) for σ− collection at $B$ = 15 T. **c,** PL peak intensity as a function of n$_{2D}$ at $B$ = 0 T (extracted from Figure 1a). At $n^*_{2D}$, the spin-triplet intensity overtakes the spin-singlet intensity due to band filling. Left (right) inset: band filling before (after) $X_2^-$ is brightened. **d,** PL peak intensity as a function of n$_{2D}$ for $B$ = 15 T. The transition to $X_2^-$ occurs at $\Delta n_{2D}$ = 3.4 × 10$^{11}$ cm$^{-2}$ earlier for σ+ collection due to the earlier filling of the +K valley at positive magnetic field. The black dotted lines correspond to the singlet-triplet crossover.

Fermi energy reaches CB2, the +K valley is reached before the −K valley, resulting in a higher σ+ emission intensity, consistent with our experimental results. Second, comparing the $X_1^-$ to $X_2^-$ transition for σ+ and σ− emission, we see the transition for σ+ occurs at a density $\Delta n_{2D} \sim 3.4 \times 10^{11}$ cm$^{-2}$ lower than for σ−, corresponding to a Zeeman splitting of 1.6 meV. This value is also consistent with the expected Zeeman spin splitting at $B$ = 15 T of $\Delta E = 2\mu_B B$ = 1.7 meV, further supporting $X_2^-$ emission with sufficient band filling. Finally, the difference in PL intensity for $X_2^-$ and $X_1^-$ is larger for σ+ than for σ−, allowing us to enhance the ratio of $X_2^-$ to $X_1^-$ PL emission under the magnetic field.

Our capability of gate tuning to access the higher conduction band with opposite spin allows us to create charged excitons with singlet and triplet spin configurations and opposite chiral light coupling. By combining long lifetime with local gate engineering[36], bright triplet charged interlayer excitons in vdW heterostructures provides a new scheme to control chiral, valley, and spin quantum states and can pave the way towards the realization of electrically controlled valleytronic devices with multiple quantum degrees of freedom.

**Acknowledgements** We thank Shiang Fang for helpful discussions. This work is supported by the DoD Vannevar Bush Faculty Fellowship (N00014-18-1-2877 for P.K., N00014-16-1-2825 for H.P.), AFOSR MURI (FA9550-17-1-0002), NSF and CUA (PHY- 1506284 and PHY-1125846 for H.P. and M.D.L.), ARL (W911NF1520067 for H.P. and M.D.L.), and Samsung Electronics (for P.K. and H.P.). Z.L and D.S. acknowledge support from the US Department of Energy (grant no. DE-FG02- 07ER46451) for magneto-photoluminescence measurements performed at the National High Magnetic Field Laboratory, which is supported by National Science Foundation through NSF/DMR-1157490, DMR-1644779 and the State of Florida. K.W. and T.T. acknowledge support from the Elemental Strategy Initiative conducted by the MEXT, Japan and the CREST (JPMJCR15F3), JST.